\documentclass[a4paper,fleqn,usenatbib]{mnras}
\usepackage{newtxtext,newtxmath}
\usepackage[T1]{fontenc}
\usepackage{ae,aecompl}	
\usepackage{graphicx,graphics}
\usepackage[flushleft]{threeparttable}
\usepackage{booktabs,caption}

\newcommand{\csec}{$\text{count}\,\text{s}^{-1}\,$}

\newcommand{\unitfl}{$\text{erg}/\text{s}/\text{cm}^{2}\,$}
\newcommand{\unilum}{$\text{erg}/\text{s}\,$}

\title[eRASSUJ 052914.9-662446]{Spectral and Timing properties of the recently discovered Be/X-ray pulsar eRASSUJ 052914.9-662446}

\author[Rai \textit{et al.}]{Binay Rai,$^{1}$\thanks{binayrai21@gmail.com}
Manoj Ghising,$^{1}$\thanks{manojghising26@gmail.com}
Mohammed Tobrej,$^{1}$\thanks{tabrez.md565@gmail.com} 
Ruchi Tamang,$^{1}$\thanks{ruchitamang76@gmail.com}
\newauthor
Bikash Chandra Paul$^{1}$\thanks{bcpaul@associates.iucaa.in}
\\
$^{1}$Department of Physics,North Bengal University,Siliguri, Darjeeling, WB, 734013, India
}

\date{Accepted XXX. Received YYY; in original form ZZZ}

\pubyear{2020}

\begin{document}
\label{firstpage}
\pagerange{\pageref{firstpage}--\pageref{lastpage}}
\maketitle

\begin{abstract}
We have presented  \emph{NuSTAR} and \emph{Swift} observations of the newly discovered Be/X-ray
pulsar eRASSU J052914.9-662446. This is the first detailed study of the temporal and spectral properties of the pulsar using 2020 observations. A coherent pulsation of 1411.5$\pm$0.5 s was detected from the source. The pulse profile was found to resemble a simple single peaked feature which may be due to emission from the surface of the neutron star only. Pulse profiles are highly energy dependent. The variation of the pulse fraction of the pulse profiles are found to be non-monotonic with energy. The 0.5-20 keV \emph{Swift} and \emph{NuSTAR} simultaneous can be fitted well with power-law modified by high energy cutoff of $\sim$ 5.7 keV. The \emph{NuSTAR} luminosity in the 0.5-79 keV energy range was $\sim$\;7.9$\times$10$^{35}$\unilum. The spectral flux in 3-79 keV shows modulation with the pulse phase.
\end{abstract}

%%%%%%%%%%%%%%%%%%%%%%%%%%%%%%%%%%%%%%%%%%%%%%%%%%

%%%%%%%%%%%%%%%%% BODY OF PAPER %%%%%%%%%%%%%%%%%%

\section{Introduction}
Neutron star X-ray binaries (NS XRBs) are categorized into two classes - High mass X-ray binaries (HMXBs) and Low mass X-ray binaries (LMXBs). Be/X-ray binaries (BeXRBs) belong to a sub-class of the HMXBs which are binary systems of a neutron star (NS) and a Be-star. A detailed information about the systems can be found in \cite{2011Ap&SS.332....1R}. However,  the majority of the X-ray pulsars discovered so far belong to the above class of binary systems. The strength of the magnetic field of the neutron star in this  system is about 10$^{12}$ G or even higher \citep{2015MNRAS.454.3760I}.
BeXRBs are mostly observed during an X-ray outburst when the X-ray flux coming  out from them is sufficient enough so that the flux
is detected by the X-ray detectors. There are two types of outbursts  namely, - Type I \& Type II. The type I outbursts  occur frequently with luminousity ($L_{x}\;\sim\;10^{36}$ \unilum) and generally originate during periastron passage when the NS passes through the circumstellar disc of the Be-star. It is also found to depend on the binary orbital phase. Type II outbursts are characterized by a significantly large luminosity ($L_{x}\;\ge\;10^{38}$ \unilum) and it lasts for the time duration of a few orbital periods. It is known  that type II outbursts may be due to warped Be-disk \citep{Okazaki2013OriginOT} and are very rarerly observed. Only a few type II outbursts are observed in a year. Persistent BeXRBs are also observed  \citep{2011Ap&SS.332....1R} but they are associated with a low luminosity ($L_{x}\lesssim10^{35}$ \unilum, less X-ray variability and  slow rotational period ($P_{s}>200$ s).
\cite{1984A&A...141...91C} observed correlation between spin period ($P_{s}$) and the orbital period ($P_{orb}$) of the NS of BeXRBs. However, with the increase in the number of BeXRBs discovery, the correlation is found to diminish significantly \citep{2016A&A...586A..81H},  providing an insight  for the better understanding of the sources. A bimodal distribution of the spin period of Be/X-ray pulsars are also reported \citep{2011Natur.479..372K}. A systematic study of 16 Be/X-ray pulsars during their quiescent states by \citep{10.1093/mnras/stx1255} reported that the source can be divided into two separate categories - (a) bright sources having hard power law spectra with luminosity about $\sim$ 10$^{34}$ erg s$^{-1}$, and (b) faint sources having thermal spectra. 
 The X-ray sources belonging to group (a) show pulsation. \citep{2013A&A...551A...1R} and are found to follow two different branches in the hardness-intensity diagram during giant outbursts in Be/X-ray pulsars. It reveals a horizontal branch during the low-intensity state and a diagonal branch during the bright intensity state with luminosity exceeding the critical luminosity. 
The recently discovered  Be/X-ray source, eRASSU J052914.9-662446 is the second Be/X-ray pulsar \citep{2020ATel13610....1M} discovered in the LMC after eRASSU J050810.4-660653 \citep{2021ATel15133....1H,10.1093/mnras/stac1820}. It was discovered during the first all sky survey (eRASS1) by \emph{eROSITA} onboard detector in the Russian/German Spektrum-Roentgen-Gamma \emph{SRG} mission. The optical spectrum of the source observed by \emph{SALT} telescope was found to be dominated by the H-alpha emission spectra, which confirms that the source  belongs to  Be/X-ray binary \citep{2020ATel13610....1M}. The OGLE I-band light curve of the optical companion shows a coherent periodicity of $\sim$ 151 days, which may be the binary orbital period of the system\citep{2020ATel13610....1M}. The pulse period of the source  was reported as $\sim$ 1412 s \citep{2020ATel13650....1M} using Nuclear Spectroscopic Telescope Array \emph{NuSTAR} observations.

\section{Observation and data reduction}
\section*{\emph{NuSTAR}}
 \emph{NuSTAR}  is a first of its kind  hard X-ray focusing telescope that operates in 3-79 keV energy range \citep{Harrison_2013}. It has 58$^{\arcsec}$ half-power diameter angular resolution. It has an energy resolution of 0.4 keV at 6 keV and temporal resolution of 0.1 ms. It consists of two identical co-aligned X-ray focusing telescopes which focus X-rays onto two independent Focal Plane Modules  namely, (i) FPMA and  (ii) FPMB. The standard processing of the data was done using NuSTARDAS software v2.1.2 and \emph{NuSTAR} Calibration Database (CALDB) v20220510 in the High Energy Astrophysics Software (HEASOFT) version 6.30.1 for spectral analysis.
The cleaned event files have been generated with the routine  \textsf{nupipeline} help file. The lightcurves, spectra including the response matrix files (RMFs) and the corresponding ancillary response files (ARFs) have been extracted using the mission specific tool , \textsc{nuproducts}. The light curves are corrected from the background using the 
\textit{ftool} \textsc{lcmath}. Light curves from the two instruments are combined using the same tool. Finally the barycentric correction was done using \textsc{barycorr} and the \emph{NuSTAR} orbit file. We consider a circular region of 25$^{\arcsec}$ radius centered around the source for a suitable extraction region of the source. A background extraction region of the same radius in the source free region is considered in our analysis. The observation ID of the \emph{NuSTAR} observation under consideration is 90601312002.

\begin{table}
 \begin{center}
 \begin{tabular}{clllc}
    \hline
    \hline
    
   Observatory	& Date of observation &	OBs ID	&	Exposure	\\
	&		&	& (in ksec)	\\
\hline	
\hline
NuSTAR		&	2020-04-08		&	90601312002	&	62.74	\\			
Swift	  &	2020-03-30   & 00013298001	  &	1.17	\\
    	   &	2020-04-08 		&   00013298003		&	0.97	\\
          & 2020-04-09      & 00013298004      &       1.68 \\

      \hline
      \hline
  \end{tabular}
  \caption{ Observation details of the source eRASSUJ 052914.9-662446.}
  \end{center}
  \end{table}

\subsection*{\emph{Swift}} 
The Neil Gehrels Swift observatory (\emph{Swift}) is a multiwavelength observatory dedicated for the study of the gamma ray burst (GRB) \citep{2004ApJ...611.1005G}. The detectors onboard \emph{Swift} are  BAT, UVOT and XRT. The X-ray Telescope (XRT) consists of a X-ray focussing Wolter I telescope and e2v CCD-22 detector \citep{2005SSRv..120..165B}. It operates in a $0.2-10$ keV energy range with  energy resolution of 140 eV at  the energy $\sim$ 5.9 keV. It has a sensitivity of $2\times10^{-14}$ \unitfl in $10^{4}$ s.  Three \emph{Swift} observations of the source have been made publicly available in the NASA data archive. The processing of the \emph{Swift}-XRT data is done using Swift Software v5.6 and \emph{Swift} CALDB v20210915. The standard screening and filtering of the Swift-XRT data are done using the task file \textsf{xrtpipeline}. We have used the \emph{Swift}-XRT event files in the photon counting (PC) in our study. The RMFs required for the spectral fitting are sourced from the latest CALDB files. A circular region of 20$^{\arcsec}$ centered around the source is considered as the source extraction region in the analysis. Another region of the same radius in the source free region was used as the background region. The source and background spectra were then  generated using these two region files. The ancillary response files (ARFs) are generated using the tool \textsf{xrtmkarf} for spectral fitting. The date, exposure and observation IDs of \emph{NuSTAR} and \emph{Swift} observations are given in Table 1.

\section{Results}
\subsection{X-ray pulsation}
The source eRASSUJ 052914.9-662446 was observed by \emph{NuSTAR} during a faint state. The X-ray variability of the source has been investigated using the light curve of 1 s binning in the energy range 3-79 keV. The average count rate of the source in the above energy range is $\sim$ 0.122\;$\pm$\;0.003 \csec. The effective exposure time of \emph{NuSTAR} on the source was $\sim$ 62.738 ks.  
We examined the signal pulsation in the light curve of the pulsar  eRASSUJ 052914.9-662446 using the methodology of the fast Fourier transformation (FFT). The FFT analysis was carried out using the \textit{ftool} \textsf{powspec}. The FFT of the source light curve is shown in the left of Figure 1. A Leahy normalized \citep{1983ApJ...266..160L} power spectrum consisting of 32768 bins in the frequency range $\sim 0.159\times10^{-4}-0.125$ Hz was generated. In a Leahy normalized power spectrum, the Poissonian noise present in the spectrum follows the $\chi^{2}$ probability distribution with two degrees of freedom (dof). The peak corresponding to the coherent pulsation was found at a frequency $\sim$ 7.08$\times$10$^{-4}$ Hz in the power spectrum. We do not find the existence of any harmonic peak in the spectrum. Using the probability distribution of the Poisson noise and the number of trials, we found that the detection of the peak is significant by more than 5$\sigma$ level \citep{2018A&A...613A..52R,Fornasini_2017exm}. The corresponding time period estimated here is 1412.4 s. 
\begin{figure*}
\begin{minipage}{0.35\textwidth}
\includegraphics[height=1.3\columnwidth, angle=-90]{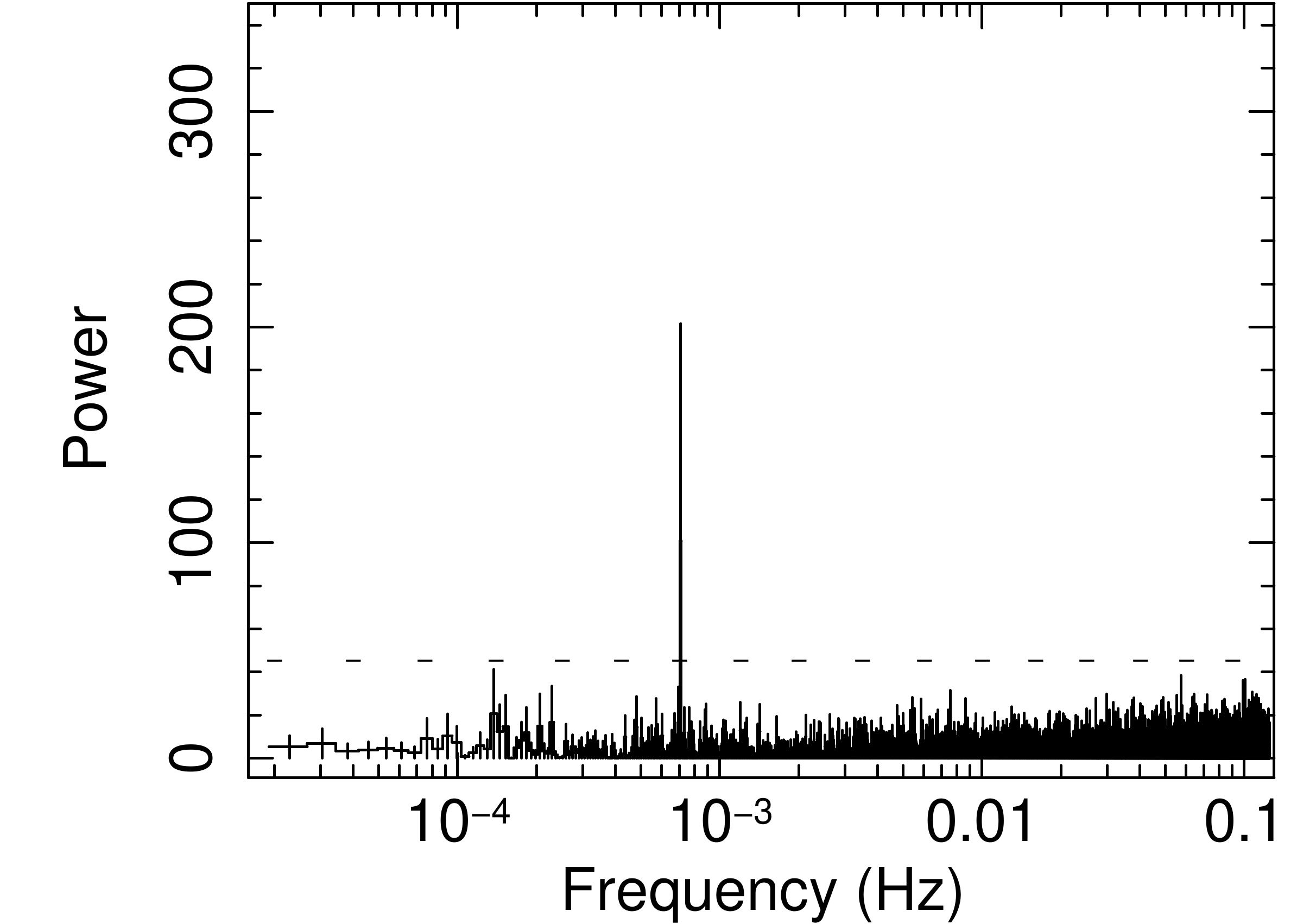}
\end{minipage}
\hspace{0.1\linewidth}
\begin{minipage}{0.35\textwidth}
\includegraphics[height=1.3\columnwidth, angle=-90]{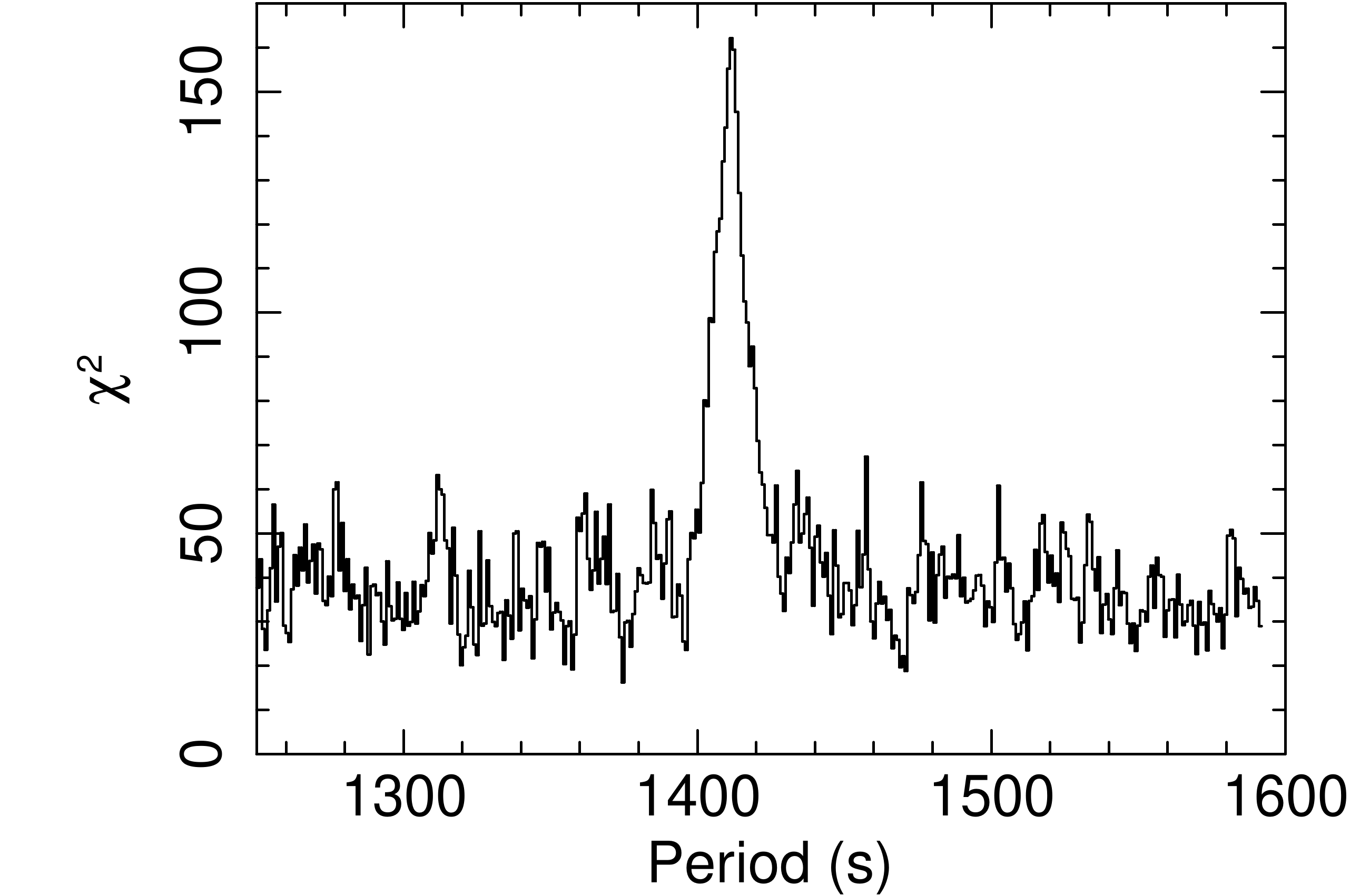}
\end{minipage}
\caption{\textit{Left - } FFT of the \emph{NuSTAR} light curve of the source. The x-axis has been plotted in log scale.A shape peak in the figure corresponding to the pulse period. The dash line indicates 5$\sigma$ significance level. \textit{Right - } Variation of $\chi^{2}$ with pulse period. The maximum value of $\chi^{2}$ corresponds to 1411.5 s.}
\label{fig-2}
\end{figure*}

The FFT method is invalid if the signal's profile is nonsinusoidal or the noise is nongaussian \citep{1996ApJ...473.1059G}. Also the FFT is best for the continuous data. We further refined the estimated pulse period of the pulsar using the epoch-folding technique \citep{1983ApJ...266..160L}. For a given trial period,  ($P_{trial}$) we determined the value of  $\chi^{2}$. If there is a presence of pulsation in the light curve then a peak is found to appear in the $\chi^{2}$ vs $P_{trial}$ plot (Figure 1). The \textit{ftool} \textsf{efsearch} is used to obtain the pulse period through this method. The best value of the pulse period for the source was estimated to be $\sim$ 1411.5 s \citep{1987A&A...180..275L}.  The advantage of the epoch folding method over the FFT is that it is independent of the shape of the light curves or signal. The error associated with the pulse period determination can be estimated by the following method given by \cite{Lut2012}. Applying  this method, we have simulated 500 light curves using the errors of the original data points. The best period of each simulated light curve is then estimated using the epoch folding method. The standard deviation of the best periods distribution is then determined, which finally gives us an estimation of the uncertainty in the pulse period. The estimated uncertainty in the pulse period is 0.5 s. 
Using the estimated pulse period, we folded the light curve to obtain the pulse profile of the pulsar. The pulse profile in the 3-79 keV resembles a simple  single-peaked behaviour. The dependence of pulse profiles on energy is explored by generating the pulse profiles in the three different energy ranges ,i.e. 3-10 keV, 10-20 keV, and 20-40 keV. The pulse profiles presented in Figure 2 reveals  its energy dependence. The pulse profile in the 3-10 keV energy range is found to resemble a  similar morphological pattern to that  in  the full energy range of 3-79 keV. However, change in its morphology is evident above 10 keV energy and one can also notice that there is a shift in the pulse phase of the corresponding maximum value of the intensity. The change in the morphology of the pulse profile with the increase in energy is common in X-ray pulsar but in this particular case it can be due to low count rate in hard energy range.
  
\begin{figure}
\includegraphics[height=9cm, width=8 cm]{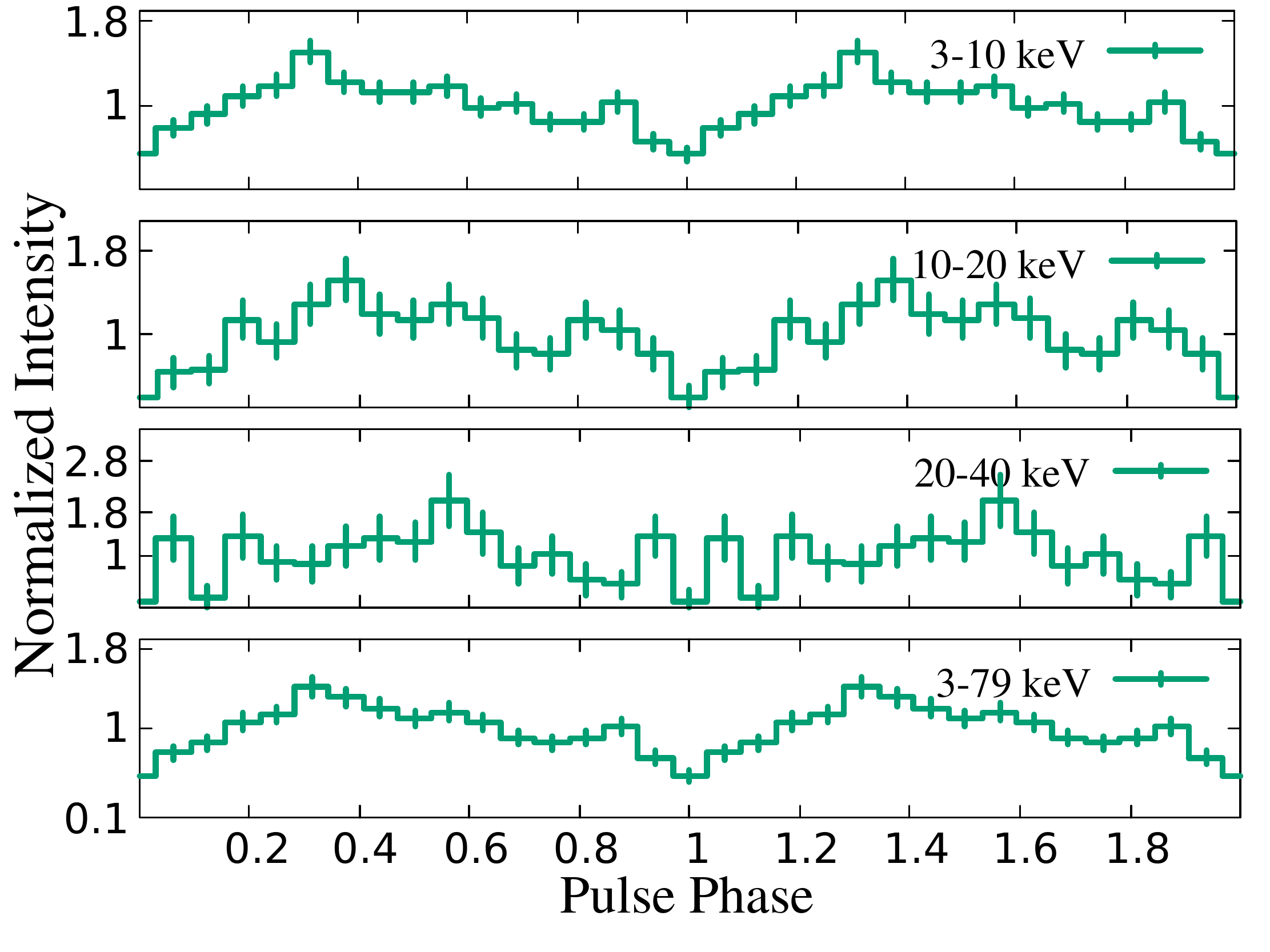}
\caption{Pulse profiles of the pulsar in three different energy bands - 3-10 keV, 10-20 keV, 20-40 keV and 3-79 keV from top to bottom, normalized at average count rates.}
\end{figure}
We also study the variation of the pulse fraction with the energy. The standard definition of the pulse fraction $(PF)$ is given by
\begin{equation}
PF=\dfrac{p_{max}-p_{min}}{p_{max}+p_{min}}
\end{equation} 
where $p_{max}$ and $p_{min}$ are the maximum and minimum intensities of the pulse profile. The variation of the pulse profile given by Eq. (1) is represented by the purple color figure in Figure 3.  We consider another definition of the pulse fraction defined as,
\begin{equation}
PF_{rms}=\dfrac{\left(\dfrac{1}{N}\Sigma^{N}_{i=1}(p_{i}-<p>)^{2}\right)^{\dfrac{1}{2}}}{<p>}
\end{equation}
where $p_{i}$ is the intensity of $i^{th}$ bin of the pulse profile, $<p>$ is the average intensity and N denotes the number of the phase bins. The r.m.s. pulse fraction is found to be lower than the pulse fraction obtained using the definition of Eq. (1). However, the variation of the pulse fraction with the energy obtained using the two different definitions given above are similar (see Fig. (4)). 
\begin{figure}
\centering
\includegraphics[scale=0.33]{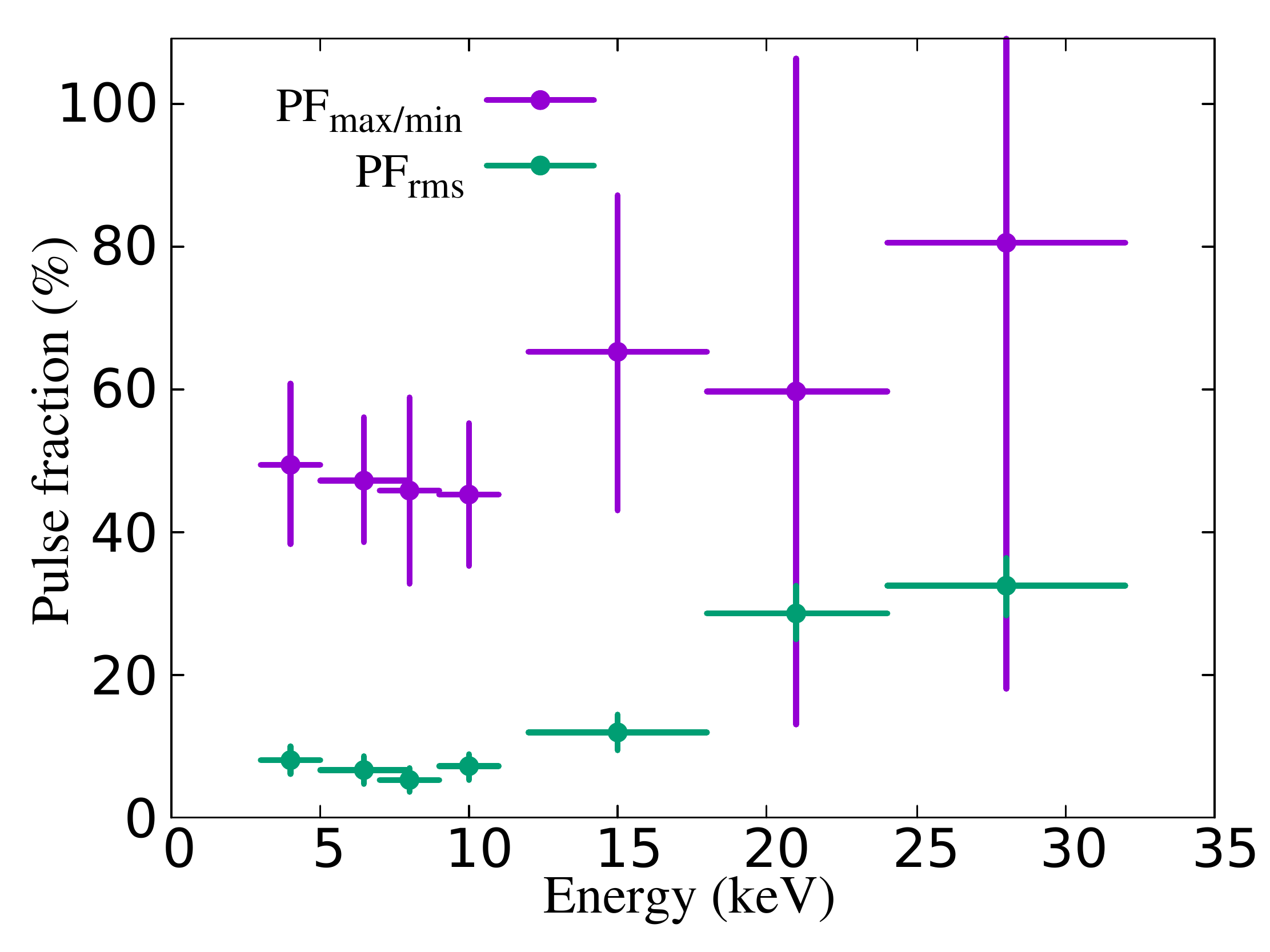}
\caption{Variation of pulse fraction with energy. $PF_{max/min}$ and $PF_{rms}$ are the pulse fraction given by Eq. 1 and 2 respectively.}
\end{figure}
The pulse fraction (PF) given by Eq. (1) is found to be 49.533 $\%$ at 4 keV. Thereafter, it  decreases slowly and attains a minimum value of 45.236 $\%$ at 10 keV. Above 10 keV, an increase in the  PF is observed. The variation of the r.m.s. pulse fraction (PF$_{rms}$) obtained  here  is also the same as that admitted by PF but the magnitude of the former diminishes from 8.162 $\%$ at  4keV to 5.297 $\%$ at 8 keV followed by an increase in its value.
\subsection{Spectral analysis}
\subsubsection*{Swift-XRT and NuSTAR simultaneous spectral fitting}
Since the \emph{NuSTAR} and one of the \emph{Swift} observation having Obs ID 00013298003 are simultaneous, we fitted \emph{Swift}-XRT and \emph{NuSTAR} spectra simultaneously. The \emph{Swift}-XRT and \emph{NuSTAR} - FPMA \& B spectra were simultaneously fitted in the energy range 0.5-20 keV, as the \emph{NuSTAR} spectra were dominated by the background above 20 keV. The spectral fitting is done in X-ray Spectral Fitting Software (XSPEC) \citep{1996ASPC..101...17A}. We have grouped the \emph{Swift} and \emph{NuSTAR} spectra such that each bin contains minimum of 1 count and followed C-statistic \citep{1979ApJ...228..939C} . The spectra were best fitted by a power-law with high energy cutoff (\textsf{cutoffpl}) model. For the estimation of the photoelectric absorption along the direction of the source, we have used \textsf{tbabs} model. The model \textsf{tbabs} was implemented using the cross-section \textsf{vern} \citep{1996ApJ...465..487V} and abundance \textsf{wilm} \citep{2000ApJ...542..914W}. The best fitted spectral parameters are shown in Table 2. The estimated column density along the direction of the was found to be $\sim$ 0.14$\times$10$^{22}$\;cm$^{-2}$ which is higher than the expected value of the column density \footnote{\url{https://heasarc.gsfc.nasa.gov/cgi-bin/Tools/w3nh/w3nh.pl?}} $\sim$6.76$\times$10$^{20}$\;cm$^{-2}$. The estimated flux in the 0.5-79 keV energy range is about $\sim$ 2.6$\times$10$^{-12}$\unitfl. Considering a distance to the source to be 50 kpc \citep{2022arXiv220300625H}, the luminosity in the 0.5-79 keV energy range was computed to be $\sim$\;7.9$\times$10$^{35}$\unilum. The fitted spectrum of the source is shown in Figure 4.
  
\begin{table}
\scalebox{0.9}{
\begin{tabular}{c c c}
\hline                                                          
Model   &       Parameters              &       Values                  \\
\hline
 & \textsf{constant*tbabs*cutoffpl} & \\
\hline                                                          
\textsf{constant}       &       $C_{FPMA}$              &       1       (fixed)         \\
        &       $C_{FPMB}$              &       1.06   $\pm$   0.05   \\
        &       $C_{XRT}$ &                0.8$\pm$0.3                     \\
\textsf{tbabs}  &       nH      ($10^{22}$ cm$^{-2}$)   &       0.1$^{+1.1}_{-0.1}$       \\
\textsf{cutoffpl}       &       photon-index ($\alpha$)         &       0.4$\pm$0.3    \\
        &       highecut        (keV )& 5.7$^{+2.3}_{-1.3}$       \\
\hline                                                          
        &       flux    ($10^{-12} erg cm^{-2} s^{-1}$) &       2.6   $\pm$   0.7   \\
                                                                
        &       $C$/dof          &       550.95/667                 \\
\hline                                                                   
\end{tabular}}
\caption{Best fitted values of the spectral parameters of \emph{NuSTAR} spectra (FPMA \& B). The errors quoted in the table are within 90$\%$ confidence interval. The flux is estimated in 0.5-79 keV energy range.}
\end{table}

\begin{figure}
\includegraphics[height=8 cm,width=6 cm,angle=-90]{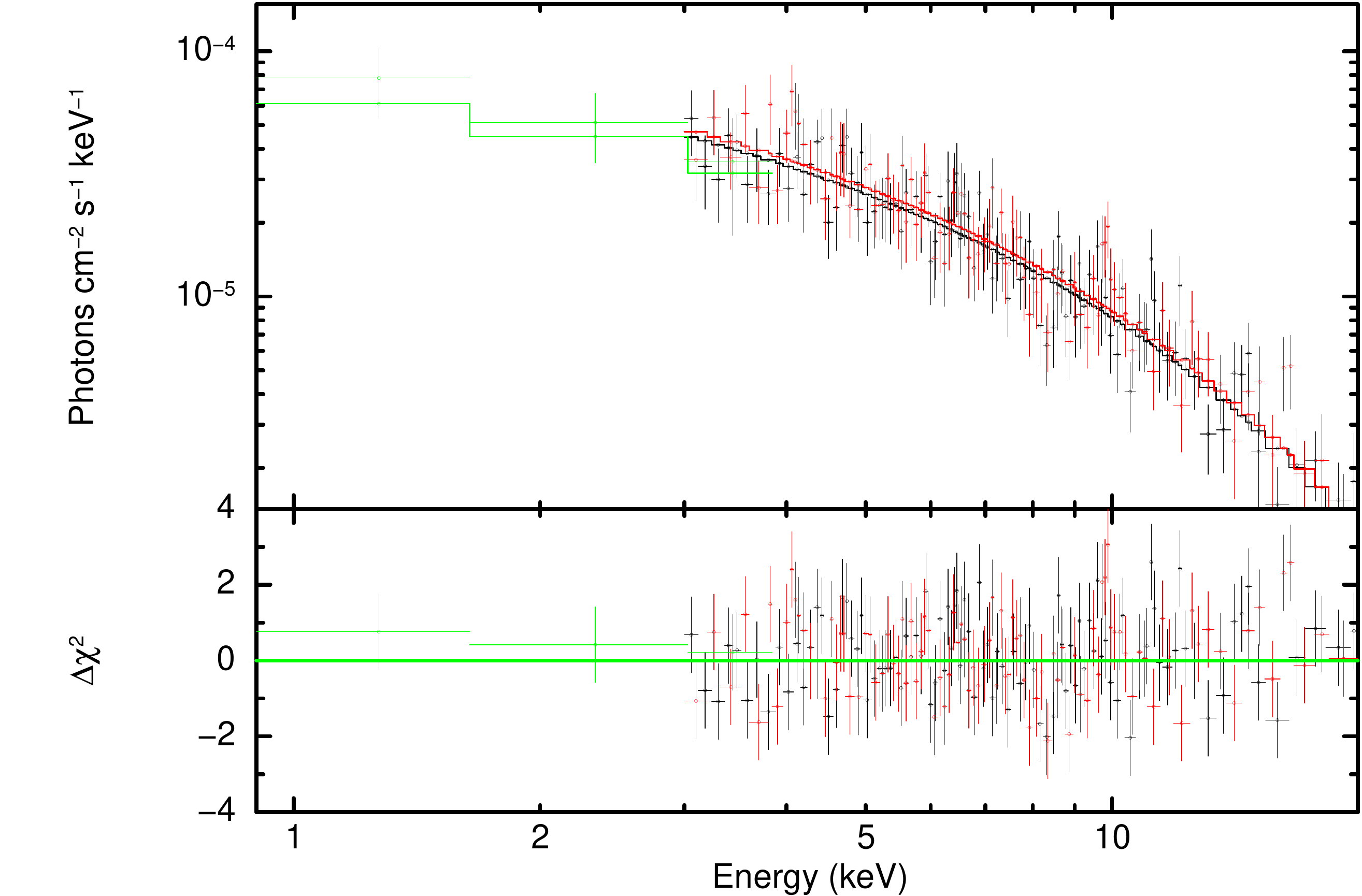}
\caption{\textit{Top panel} - The unfolded spectra of the pulsar, the green, black and red points represent \emph{Swift}-XRT, FPMA and FPMB spectra respectively. The green, black and red lines represent best fitted lines for the \emph{Swift}-XRT, FPMA \& B spectra respectively. \textit{Bottom panel} - Residuals left after fitting.}
\end{figure}
\subsubsection*{\emph{Swift-XRT} spectral fitting}
The three \emph{Swift}-XRT spectra in the 0.5-10 keV energy range were fitted using \textsf{tbabs} and \textsf{power-law} models. The spectra were grouped in such a way that each bin contains a minimum of 1 count. We adopted C-statistics while fitting the spectra \footnote{\url{https://www.swift.ac.uk/analysis/xrt/spectra.php}}. We were unable to constrain the value of column density (nH) for the ObsID 00013298001 and 00013298004, so we fixed it to the expected value of 6.76$\times$10$^{20}$ cm$^{2}$. The main motivation behind fitting the \emph{Swift}-XRT spectra is to estimate the flux (see Table 3) of the source.

\begin{table*}
\centering
\scalebox{1}{
\begin{tabular}{cccc}
\hline
ObsID	&	00013298001		&		00013298003		&	00013298004 \\
\hline		
time (MJD) & 58938.57		&		58947.067		&	58948.063\\
nH ($10^{20}$ cm$^{-2}$)	&	6.76	(fixed)		&	6.76 (fixed)
	&	6.76(fixed) \\		
photon-index	&	-0.479	$^{+2.5}_{-0.5}$	&	1.0$\pm$1.3	&	0.4$\pm$1.1\\
flux ($10^{-12} erg cm^{-2} s^{-1}$)	&	3.4$^{+3.1}_{-3.3}$	&	1.1$^{+1.4}_{-0.7}$	&	1.7$^{+3.2}_{-1.2}$\\
C/dof	&	15.52/14	&	10.51/21	&	15.73/21\\
\hline
\end{tabular}}
\caption{The fitted parameters \emph{Swift}-XRT spectra. The models used were \textsf{tbabs*powerlaw}. C denotes the fitting statistics. The errors quoted above are within 90$\%$ confidence range. The flux is estimated in 0.5-10 keV energy range.}
\end{table*}

\subsubsection*{Phase-resolved analysis}
Phase-resolved spectroscopy provides us with the diagnostic tool to study the geometry of the emission region close to the surface of the neutron star by looking at the variation of spectral parameters with the pulse phase. So for this purpose, we have extracted the source spectra for 10 different phase intervals each of size 0.1. The spectra were fitted with the \textsf{cutoffpl} model. We have used \textsf{tbabs} model for the estimation of photoelectric absorption along the direction of the source.  We were unable to constraint the value of nH and the cutoff energy of the \textsc{cutoffpl}. Therfore we fixed the nH value to its expected value of $\sim$ 6.76$\times$10$^{20}$ cm$^{-2}$ and the cutoffpl energy to the value obtained from the phase-averaged spectral fitting. The spectra were grouped in such a way that each bin contains a minimum of 1 count. We have adopted C-statistics to obtain the fit statistics in these cases. The variation of the spectral parameters with the pulse phase is shown in Figure 5. Due to large uncertainties associated with the photon index it is hard to say that photon-index shows a modulation with pulse phases (Figure 5)  but the flux shows some modulation with the pulse phase. 
 
\begin{figure}
\centering
\includegraphics[width=8 cm,height=7cm]{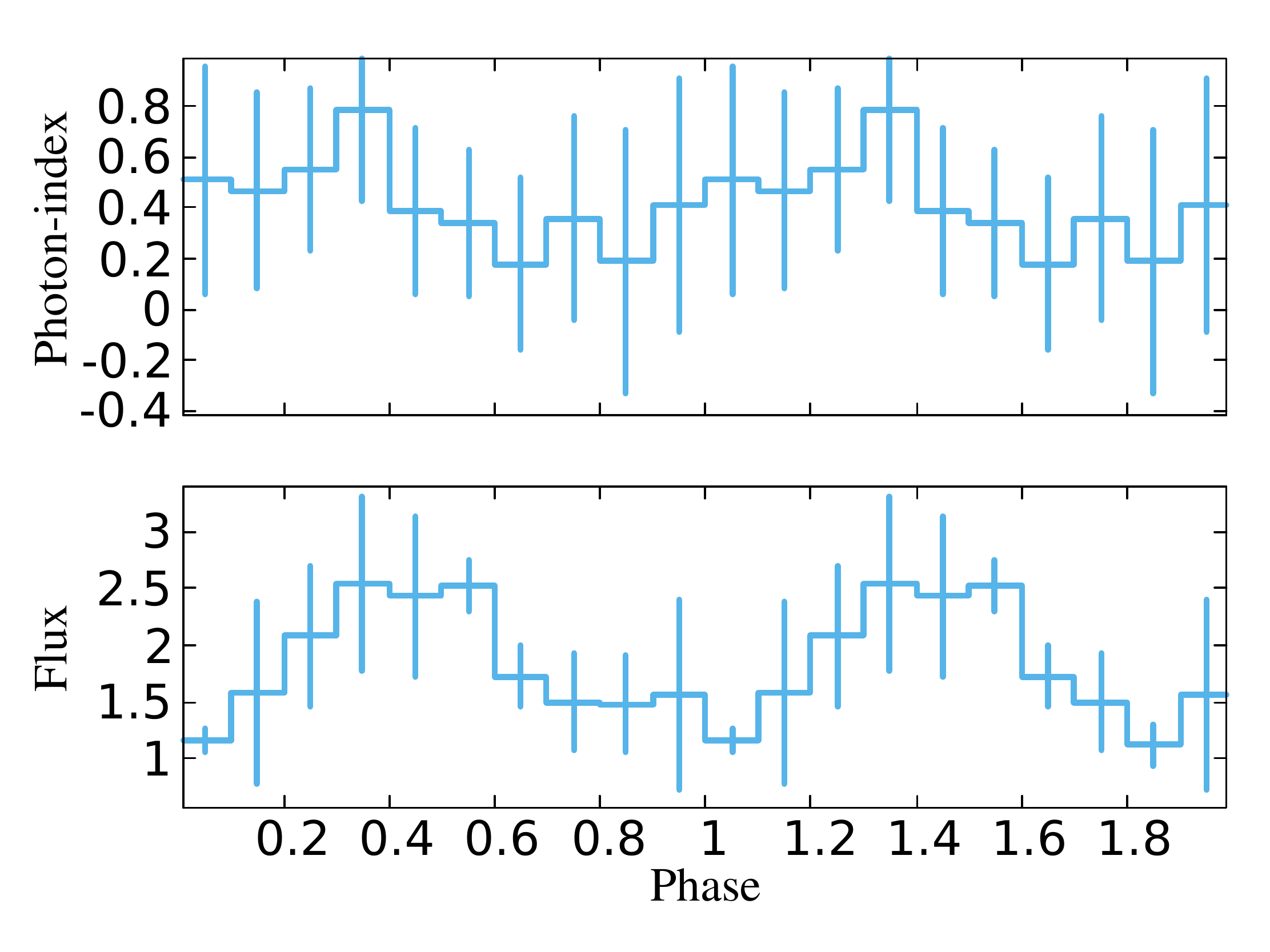}
\caption{Variation of photon-index (top) and flux (bottom). The flux is on the scale of 10$^{-12}$ \unitfl and is measured in 3-79 keV energy range.}
\end{figure}

\section{Discussion}
In the present paper we consider a study of recently  discovered Be/X-ray pulsar namely,  eRASSUJ 052914.9-662446 which is a slowly rotating X-ray pulsar ($P_{s}>200\;s$) and  has a pulse period of $\sim$ 1411.5$\pm$0.5 s. This pulse period is not orbital corrected as the orbital parameters of the source is not known. The other Be/X-ray pulsars with pulse period greater than 1000 s are SXP1062 \citep{2012MNRAS.420L..13H}, SXP1323 \citep{2000A&AS..147...75S} and SXP4693 \citep{2010ApJ...716.1217L}. SXP1062 and SXP1323 are the X-ray pulsars that show largest observed spin-up rate \citep{2017A&A...602A..81C,2020A&A...637A..33T}. These two sources  are  associated with the supernova remnant \citep{2012MNRAS.420L..13H,2019MNRAS.485L...6G}. The \emph{Swift}-XRT flux in the (0.5-10) keV energy range is found to lie between $(1.1-3.4)\times10^{-12}$ \unitfl. The \emph{Swift}-XRT flux decreases from $3.4\times10^{-12}$ \unitfl to $1.7\times10^{-12}$ \unitfl in $\sim$ 9.5 days. Assume a distance of the source to 50 kpc,  the luminosity of the source in the (0.5-10 keV) energy interval is found to decrease from $\sim 1\times10^{36}$ \unilum to $5.1\times10^{35}$ \unilum. So we do not find any abrupt change in the source luminosity in between the given time interval. The small spectral variability of the source indicates that  it might be of persistent nature, however the luminosity is found 5-10 times more than the persistent BeXBs ($L_{x}\lesssim10^{35}$ \unilum,\cite{2011Ap&SS.332....1R}). The 0.5-20 keV spectra of the pulsar is fitted well with a powerlaw model of photon index 0.4 modified by cutoff energy of 5.7 keV. The discrepancy in the value of the photon index and cutoff energy obtained here as compared to \cite{2022arXiv220901664M} can be due to the difference in the model used for the estimation of the column density. \citep{2022arXiv220901664M} have used partial covering \textsc{tbpcf} model to estimate the column density where as we have used \textsc{tbabs} model to estimate it value. The spectral fitting performed by \cite{2022arXiv220901664M} was in 0.2-78 keV whereas we have performed spectral fitting in 0.5-20 keV energy range. However, the flux estimated here is almost same as that compared to \cite{2022arXiv220901664M}. The flux measured in 0.5-79 keV energy range $\sim$ 2.6$\times$10$^{-12}$\unitfl, which is almost two times greater than the \emph{Swift}-XRT flux in 0.5-10 keV energy range. Absence of iron emission line at 6.4 keV in \emph{NuSTAR} spectra of the source indicates an absence of disc or may be a  very small disc in case it exists. The emission  line is due to reprocessing of X-ray due to cool matter present in the disc and commonly found in BeXBs \citep{2013A&A...551A...1R}. The detection of iron line provides direct evidence of the presence of matter in the surroundings of the neutron star. A presence of cyclotron line in the spectra of X-ray pulsars provides a direct estimate of the magnetic field of the pulsars. However in absence of the cyclotron line the spin-up or spin-down rates of the pulsars are used to estimated the magnetic field of the pulsars. We did not found an evidence of cyclotron line in the 0.5-20 keV spectrum of the pulsar. As the \emph{Swift}-XRT observations were of very short durations, we cannot use these observations to estimated the pulse period. Due to scarcity of the data it is is not possible to constrain the magnetic field of the pulsar directly. So we can use indirect method to estimated the magnetic field of the pulsar. One such method is based on the quasi-spherical accretion model in X-ray pulsar given by \cite{2012MNRAS.420..216S}. The low luminosity (of order 10$^{35}$ \unilum) and long pulse period (> 200 s) of the pulsar supports quasi-spherical accretion from stellar wind in the source \citep{10.1093/mnras/stx3127,2015MNRAS.446.1013P,2022arXiv220901664M,2020MNRAS.498.4830J}. In this accretion process, a hot quasi-spherical shell is formed around a magnetospheric boundary of the pulsar from where the mass is accreted onto the surface of the neutron star. Depending on the mass accretion rate, a pulsar will either spin-up or spin-down. Assuming spin equilibrium of the spin period in this accretion regime, the magnetic field strength can be estimated using the following equation \citep{2015MNRAS.446.1013P},

\begin{equation}
P_{eq}\approx940[s]\mu_{30}^{12/11}\left(\dfrac{P_{orb}}{10 d}\right)\dot{M}_{16}^{-4/11}v^{4}_{8},
\end{equation}
where $\mu_{30}=\mu/10^{30}$ [g cm$^{-2}$] is the neutron star (NS) dipole magnetic moment, $\dot{M}_{16}=\dot{M}/10^{16}$ [g\;s$^{-1}$] is the rate of mass accreting onto the surface of the NS, $P_{orb}$ is the orbital period of the binary system and $v_{8}=v/10^{8}$ [cm\;s$^{-1}$] is the velocity of the stellar wind. The luminosity in 3-79 keV energy range for the \emph{NuSTAR} is $\sim$7.9$\times$10$^{35}$ \unilum. The luminosity is related to the mass accretion rate as $L=0.1\dot{M}c^{2}$, where $c$ is the velocity of the light in vacuum. The mass accretion rate ($\dot{M}_{16}$) for the given luminosity is $\sim$0.9. Considering $P_{orb}$ as 151 days \citep{2022arXiv220901664M}, $P_{eq}$ as 1412 s and $v_{8}$ about 0.2, which is the typical value of wind velocities observed in Be stars \citep{1988A&A...198..200W}, we get $\mu_{30}=$59.8. This corresponds to a magnetic field of $\sim$6$\times$10$^{13}$ G. It has been found that for pulsars with spin period $P_{s} \gtrsim$ 1000 s, the estimated magnetic field is very high ($>$ 10$^{13}$ G) like in SXP 1062 \citep{2012ApJ...757..171F}, SXP 1323 \citep{2022A&A...661A..33M}, 2S 0114+650 \citep{1999ApJ...513L..45L} and 4U 2206+54 \citep{10.1111/j.1365-2966.2012.21509.x}. \cite{1999ApJ...513L..45L} to explain the long spin period ($\sim$ 2.7 hr) in 2S 0144+650 argued that its magnetic field should be $\gtrsim$ 10$^{14}$ G. Similarly \cite{10.1111/j.1365-2966.2012.21509.x} proposed that a high magnetar-like field can be produced if the $P_{s} > 1000 s$. \cite{2012A&A...540L...1D} also showed that for slow rotating pulsar the magnetic field should be $>$ 10$^{14}$ G. So it is possible that the magnetic field of the eRASSUJ 052914.9-662446 to be $>$ 10$^{13}$ G. A pulsar with magnetic field $>$ 10$^{13}$ G and $P_{s} >$ 1000 s can never reach a propeller phase if it accretes through a disc, then it is possible for the pulsar to pass through a cold accretion state \citep{2017A&A...608A..17T}.

The \emph{NuSTAR} pulse profile in 3-79 keV energy range of the source is single peaked and no additional feature like dips is observed. The luminosity in 0.5-79 keV energy range is $\sim$\;7.9$\times$10$^{35}$ \unilum, at this luminosity we cannot expect an extended accretion column to form \citep{2015MNRAS.447.1847M,10.1093/mnras/stv2087}. So most of the X-ray photons should originate from the region very close to the surface of the neutron star. In such a case the emission pattern should be of pencil beam shaped \citep{1976MNRAS.175..395B}. The simple single-peaked pulse profile in 3-79 keV energy range is supported by  argument \citep{2017MNRAS.470.4354V}. At a very high luminosity state one can expect formation of an accretion column which results in a fan-shaped emission pattern with a complex pulse profile. Actually the two different beam emission patterns linked with two different accretion regimes are well separated by a certain luminosity called critical luminosity $(L_{crit})$ \citep{1976MNRAS.175..395B}. At a luminosity above the critical luminosity the infalling X-ray emitting plasma may be stopped above the surface of the neutron star by a radiation dominated shock in the accretion column. The matter slowly gets decelerated as it moves through the shock and the X-ray emission occurs through the side wall of the accretion column, perpendicular to it,  resulting in fan-shaped beaming pattern \citep{2012A&A...544A.123B}. For luminosity well below the critical luminosity no radiation dominated shock is formed and the emission of radiation occurs from the surface of the neutron star forming a pencil beam, perpendicular to the surface of the neutron star. The shape of the pulse profiles were found to vary with the energy. The pulse fraction (PF and PF$_{rms}$) of the pulse profile is also found to depend on energy which does not change monotonically but rather it decreases with the energy initially ($\leq$ 10 keV) and thereafter it is found to increase. The pulse fraction of X-ray pulsars show local feature like maxima or minima near the cyclotron line energy \citep{2009AstL...35..433L,2010MNRAS.401.1628T,2021ApJ...915L..27M}.  For hard X-ray ($>$ 10 keV) range the pulse fraction of X-ray pulsars are found to increase with increase in energy (except presence of other local features). With the increase in energy the pulse profile becomes less structured due to which the pulse fraction is found to increase \citep{1997ApJS..113..367B,2019A&A...622A..61S}.

\section*{Data availability}

The data used in this research are downloaded from NASA HEASARC data archive.

\section*{Acknowledgements}

This work made use of \emph{Swift} and \emph{NuSTAR} data downloaded from NASA the high energy astrophysics science archive research center (HEASARC) data archive. \emph{NuSTAR} is a project led by Caltech and funded by NASA. This research also has made use of the nustar data analysis software (NUSTARDAS) jointly developed by the ASI science data center (ADSC), Italy and Caltech. This research has made use of the software provided HEASARC, which is supported by the Astrophysics Division at NASA/GSFC and the high energy astrophysics division of the Smithsonian Astrophysical Observatory (SAO). The authors would like to thank ICARD, NBU for the providing extended research facilities. The authors of this paper is grateful to the anonymous reviewer for his/her valuable suggestion.

%%%%%%%%%%%%%%%%%%%%%%%%%%%%%%%%%%%%%%%%%%%%%%%%%%



\bibliographystyle{mnras}
\bibliography{erassu.bib} % if your bibtex file is called example.bib

%\begin{thebibliography}{99}

%\bibitem[\protect\citeauthoryear{}{}]{}

%%%%%%%%%%%%%%%%%%%%%%%%%%%%%%%%%%%%%%%%%%%%%%%%%%

%%%%%%%%%%%%%%%%% APPENDICES %%%%%%%%%%%%%%%%%%%%%

%%%%%%%%%%%%%%%%%%%%%%%%%%%%%%%%%%%%%%%%%%%%%%%%%%

%\end{thebibliography}

% Don't change these lines

\bsp	% typesetting comment
\label{lastpage}
\end{document}